\documentclass{article}

\usepackage{arxiv}
\usepackage[utf8]{inputenc} 
\usepackage[T1]{fontenc}    
\usepackage{hyperref}       
\usepackage{url}            
\usepackage{booktabs}       
\usepackage{amsfonts}       
\usepackage{nicefrac}       
\usepackage{microtype}      
\usepackage{lipsum}
\usepackage{graphicx}
\usepackage{tikz}
\usepackage{array}
\usepackage{amsmath,amssymb}
\usepackage{pifont}
\newcommand{\xmark}{\ding{55}}%

\title{Are PETs (Privacy Enhancing Technologies) Giving Protection for Smartphones?- A Case Study}

\author{
  Tanusree Sharma\\
  Illinois Informatics Institute\\
  University of Illinois at Urbana Champaign\\
  Champaign, IL 61820 \\
  \texttt{tsharma6@illinois.edu} \\
   \And
 Masooda Bashir \\
  School of Information Sciences\\
  University of Illinois at Urbana Champaign\\
  Champaign, IL 61820\\
  \texttt{mnb@illinois.edu} \\
}

\begin{document}
\maketitle

\begin{abstract}
With smartphone technology’s enhanced way of interacting with the world around us, it has also been paving the way for easier access to our private and personal information. This has been amplified by the existence of numerous embedded sensors utilized by millions of apps to users. While mobile apps have positively transformed many aspects of our lives with new functionalities, many of these applications are taking advantage of vast amounts of data, privacy apps, a form of Privacy-Enhancing Technology can be an effective privacy management tool for smartphones. To protect against vulnerabilities related to the collection, storage, and sharing of sensitive data, developers are building numerous privacy apps. However, there has been a lack of discretion in this particular area which calls for a proper assessment to understand the far-reaching utilization of these apps among users. During this process we have conducted an evaluation of the most popular privacy apps from our total collection of five hundred and twelve to demonstrate their functionality specific data protections they are claiming to offer, both technologically and conventionally, measuring up to standards. Taking their offered security functionalities as a scale, we conducted forensic experiments to indicate where they are failing to be consistent in maintaining protection. For legitimate validation of security gaps in assessed privacy apps, we have also utilized NIST and OWASP guidelines. We believe this study will be efficacious for continuous improvement and can be considered as a foundation towards a common standard for privacy and security measures for an app's development stage.

\end{abstract}


\section{Introduction}

Information security and privacy is an issue concerned with the proper supervision of data consent, notice, and regulations. More specifically, practical data privacy concerns often revolve around a few common questions: "How data is legally collected or stored?” and “Whether or how data is shared with different parties?" On-going research into information security and privacy strives to pin down frequent issues related to information security violations and possible solutions \cite{smith2011information}. Recently discussed topics on use of technology such as contact tracing, quarantine tracker for combating COVID-19
 \cite{sharma2020use} are one of few most critical aspects where data handling is such a big concerns.  Data is one of the most important assets, companies and governments are concerned about. With the ever-growing amount of data in online space, companies find enormous value in collecting, sharing, processing, and using data for their own use. Big tech giants such as Google, Facebook, Twitter, Amazon, and eBay have been building their billion-dollar empires atop the data economy. In this data life-cycle, transparency in requesting consent, abiding by privacy policies, and managing and collecting data are important requirements for building trust and accountability with customers who expect privacy. While we have different ways to ensure the privacy of our physical data, we are losing control over our digital data by various means. We are frequently giving up a vast amount of sensitive information and revealing pieces of our existence on digital platforms. All online spaces we enter glean some aspect of our personal information. Different parties connect the pieces of information for mapping a very detailed personal profile and slowly manipulating us into changing our cognition on certain topics in different digital communication platform. We are aware from history that knowledge of someone’s ethnic background, political or religious beliefs can be life threatening if it falls into the wrong hands. One of the many events that several researchers have argued relates to the topic of privacy violation and autonomy is the 2016 United States Presidential Election \cite{isaak2018user}, \cite{ward2018social}. According to its study, Cambridge Analytica’s methods of data collection and commercial use represent significant violations of personal autonomy and privacy, which is stated as psychological traits and, in particular, behavioral microtargeting \cite{isaak2018user}, \cite{ward2018social}. It may also seem unusual when you and your friends pay different airfares because of the zipcode they live in. These real-life cases make it clear that our data security and privacy in online has been taken away gradually. Moreover, cyber-attacks such as, phishing email \cite{sharma2020analysis}, social engineering strategies for spear phishing \cite{krombholz2015advanced} are evolving with technological advancement which are paving the way for attackers to trap users to get into their system for getting access of their personal information. Therefore, new threat landscapes are being spread so frequently to be addressed.

The widespread usage of smartphones is allowing the data life-cycle to be used more efficiently for data collectors and leaving other parties (e.g., users, consumers) helpless. A Smartphone’s ability to create easier modes of communication via instant messaging, mobile browsers, highly functioning cameras is allowing data collectors to grasp the information of users. Mobile apps are taking this process one step forward by advanced processing capabilities to store and process sensitive information as well, such as the user’s location, lists of contacts, and personal photographs, which require adequate protection. A form of privacy-enhancing technology – privacy apps have been designed to play an important role in providing privacy-preserving environments with privacy mechanism, for example, smart encryption, secure session management, secure inter-process communication, and secure transport layers. To date, privacy apps are not explored on their privacy-preserving computation protocols \cite{huang2011privacy} and to the best of our knowledge, there is not enough analysis of privacy apps’ mechanisms and how they keep up to the privacy requirement standards in their base development process. While ensuring security and privacy involves developer industries, users, and government where each of these entities relies on others, this process requires consideration of both technology and regulations. Therefore, it is fair for any party to question if the necessary security and privacy mechanisms are in place they are supposed to or claimed to offer based on their service \cite{tanusree2020privacy}. We have encountered literature where different self-made guidelines, catalogues and practices in different app industries on their development life cycle which may create conflicts of interest and will put the privacy of users at stake \cite{sunyaev2015availability}. To put our understanding on point, we explored existing guidelines and frameworks which can provide certain requirements/criteria lists for ensuring security and privacy for mobile OS, in particularly, NIST.SP.800-163 for "Vetting the Security of Mobile Applications" \cite{quirolgico2015vetting}. NIST is a non-regulatory government agency that develops technology, metrics, and standards to drive innovation and economic competitiveness at U.S.-based organizations in the science and technology industry that assists in protecting information \cite{radack2011managing}. Compliance with NIST standards and guidelines has become a top priority in many high-tech industries. 

In our research, we have outlined our project for analysis of these “Privacy App” from forensics and policy point of view. For the part of forensic, we first utilized automated framework MobSF(Mobile Security Framework) for detecting app specific security and privacy concerns, followed by rigorous dynamic analysis to highlight vulnerabilities on information, accessibility, and transmission, purpose of requesting and processing it. We have also utilized testing tools (fiddler, SSL labs) for measuring the safety of data transmission and communication in run-time. Finally, we utilized $NIST.SP.800-163$, OWASP guidelines, Solove’s Privacy Taxonomy \cite{daniel2006taxonomy} as a mixed method approach to assign how these “Privacy apps’” modules and components are keeping up with data protection component and if they are satisfying those requirements for ensuring security and privacy. Our goal of systematic evaluation of privacy apps is to identify missing pieces from both security and privacy point of view for developing novel frameworks to help identify and mitigate the security and privacy risks  \cite{sharma2020preserving} that arise from smartphones applications in the long run.

The paper proceeds as follows. In section 2, we extensively reviewed security and privacy related research on smartphone app to indicate why there still needs to consider additional techniques for preserve privacy framework. With substantial research evidence of apps security and privacy concerns and area for future scopes, in section 3, we designed our experiment model to conduct our assessment in 3 different ways (manual, static and dynamic). According to our model of analysis, we further described our steps independently for better clarification with retrieved results in section 4.  In section 5, we took an additional approach for conferring our analyzed results with well-developed existing security and privacy guidelines (NIST, OWASP, Solve’s Taxonomy) which has been recommend to be taken under consideration by researcher in case of security and privacy for a long time now. In Section 6 discusses our findings and implications, followed by a conclusion in section 7.

\section{Background}
Apple App Store and the Google Play Store are two most important marketplaces in the world for publishing apps. The popularity of smartphones is constantly growing, with an increased demand for applications that has resulted in serious privacy concerns. Smartphones’ advanced processing capabilities are allowing most apps to store and process sensitive identifiable information. People are using different kind of apps for their daily life, even without realizing the potential negative impact. Mobile phones’ progressive adoption in people’s daily activities is shaping and defining mobile computing outburst. To get this situation under control, it is time to address apps’ privacy issues and identify primary concerns from privacy and security point of views. In previous scholarly research and published studies, we have found assessments of privacy risks and analyses of particular types of apps, for example, mHealth, fitness apps and ebanking apps, etc. Those analyzed apps are mostly from the Android marketplace. Most of them considered doing their evaluation in terms of examining if apps have satisfied the current security and privacy legislation, and standards and certifications for security and privacy \cite{papageorgiou2018security},\cite{huckvale2015unaddressed}. For our research, we have considered the privacy risks and suggested mechanisms recorded by NIST \cite{ogata2018vetting} and OWASP \footnote{OWASP Mobile Top 10. Retrieved from
https://owasp.org/www-project-mobile-top-10/
} . For our experiment, we chose iOS privacy apps to better understand the particular privacy mechanisms for certain privacy vulnerabilities which may not be present in regular apps. We explored the details about iOS application structure, Inter-process communication (IPC), iOS application publishing, and iOS application attacks from Apple developers’ documentation.

iOS is isolated in comparison to Android and has a mandatory access control (MAC) mechanism, with IPC options to minimize potential attacks. The iOS platform offers privacy and data protection advantages through uniform hardware/software integration, secure boot, hardware-backed keychain \footnote{Apple. iOS Security. Retrieved from https://www.apple.com} , and file system encryption\footnote{Apple App developers Documentation. Retrieved from https://developer.apple.com/documentation/uikit}. Despite having built-in security measures for apps, iOS app developers still need to worry about data protection because of the availability of mobile devices and the different sensors associated with apps installed on them. There is a varied amount of concern related to data protection, keychain management, Touch ID/Face ID authentication, and network security layers from functional as well as non-functional points of view. Keeping those in mind, there still needs to be security and privacy testing and reverse engineering for proper validation of security measures, including how they are working on run-time and how vulnerabilities can cause fatal privacy threats to users’ personal information. An iOS application attack surface consists of all components of the application, including the supportive material necessary to release the app and to support its functioning. During this process, iOS application may be vulnerable to attack if it does not: validate all input by means of IPC communication or URL-schemes; validate all input by the user in input fields; validate the content loaded inside a WebView; Securely communicate with backend servers to avoid being susceptible to man-in-the-middle (MITM) attacks between the server and the mobile application; securely store all local data, or load untrusted data from storage; and protect itself against compromised environments, repackaging or other local attacks. After having all the advanced technologies included as part of a built-in design model in iOS, there still remain vulnerable situations stemming from smartphones’ sensors to gaining access to users’ information.

We also found evidence of severe vulnerabilities that remained unaddressed and several new privacy issues from existing studies and different reports. According to their findings mobile apps are still vulnerable to attacks and complete security is not quite an achievable task, even when enforcing privacy and security requirements \cite{spensky2016sok}. There are many well-known popular apps that process massive amounts of sensitive data for big data analytics and yet fail to provide basic protection of users’ privacy due to their inappropriate implementation, poor design choices and lack of proper of guidelines from a validated organization (e.g., NIST, IAPP) \cite{ogata2018vetting}, \cite{rosenfeld2017data}. There is an emerging shift toward mobile applications where the goal is to make life more comfortable in respect to health service, banking, shopping, fitness, and even dating \cite{papageorgiou2018security}, \cite{akinyede2017development}, \cite{vitak2018privacy}. This noticeable growth of mobile apps comes with growing concern for users’ data privacy on their mobile devices. According to a Homeland Security study on mobile device security, threats to the Government’s use of mobile devices are real and exist across all elements of the mobile ecosystem \footnote{	Homeland Security. Study on Mobile Device Security. Retrieved from https://www.dhs.gov/}. In order to ensure some level of privacy protection for mobile phone and app users, there is another kind of app that is being developed called privacy apps – a form of privacy enhancing technology. To preserve privacy and give users data protection \footnote{Office of Privacy Commissioner of Canada. Privacy Enhancing Technologies – A Review of Tools and Techniques. Retrieved from https://www.priv.gc.ca/en/}, there are many privacy enhancing technologies already available \cite{chun2015privacy}. App developers are also designing different privacy apps to give protection to categorized data, for example, privacy apps for photos, videos, passwords, browsers, and adblocking. However, it is still questionable if those privacy apps actually protect privacy or are perhaps taking away part of users’ data, and sharing and processing it for their own benefits. From a technological point of view, other concerns can be added to the earlier one: are those privacy apps following proper guidelines at their development phase, and are those guidelines comprehensive for all of the privacy apps out there? As those are intended to ensure privacy to users, it is necessary to analyze if those apps are including adequate privacy mechanisms and how secure those apps are to use. IAPP compares different guidelines for mobile applications in terms of privacy, security, data collection, retention, notice, and consent \footnote{iapp. COMPARISON OF MOBILE APPLICATION GUIDE- LINES.Retrieved from
https://iapp.org/resources/comparison-of-mobileapplication-guidelines/
}.

In this study, we are mainly utilizing Mobile Security Framework (MobSF) which is an automated, open source pen-testing framework capable of performing static, dynamic and malware analysis. This is mostly recommended for static analysis of security in mobile applications. It can be used for effective and fast security analysis of iOS mobile applications and supports IPA binary and zipped source code. It has a web service that consists of a dashboard which presents the results of the analysis, a documentation site, an integrated emulator, and an API that allows users to trigger the analysis automatically. For dynamic analyses, we used Fiddler, an HTTP debugging proxy server application that can record data shared over HTTP and other vendors and measure security. We have also utilized SSL Labs (https://www.ssllabs.com/ssltest/) from Qualys. SSL verification is necessary to ensure certificate parameters are as expected. There are multiple ways to check the SSL certificate for our analyses. After performing the analyses, we employed NIST mobile privacy guidelines to compare and contrast with three different analyses results. We utilized NIST guidelines for mobile apps privacy and security and OWASP Mobile Security Testing Guide \cite{franklin2019mobile},\cite{mueller2017owasp} for analyzing selected privacy apps concerning privacy and security. Our goal is to provide systematic feedback about privacy mechanisms to the developers of those apps. We are analyzing privacy mechanisms of the selected apps to check if they meet the requirements of NIST guidelines and, at the same time, we are studying the comprehensiveness of our utilized mobile testing framework MobSF. Because of the ubiquitous terminology of "privacy" in our modern technology, it is challenging to address privacy issues within our existing testing framework. While we are using NIST and OWASP apps privacy guideline requirements, we are also pinpointing some of the requirements which are present in our guidelines but are not present in our experiment environment.

\section{Methodology}
In this section, we first present our overall app collection and criteria for selecting manageable counts of apps. Our experiments have been conducted on iOS privacy apps. Secondly, we describe how we conducted the assessment for investigating the security and privacy components offered by each app. We conducted our assessment with three different types of experiments: manual, static and dynamic analysis by the help of OWASP Mobile Security Testing Guide. In addition to providing a finding’s report, we performed a NIST guideline auditing procedure to determine whether the reviewed apps have the privacy mechanisms needed to combat against the current threats landscape. To accomplish the analysis, our method includes two parts: data collection and analyses (Manual analysis, Static code Analysis, dynamic/ run-time analysis). Figure 1 presents the overall apps collection with our assigned categories. During our research, experiments progress through recording the results from frameworks for each apps and comparing their mechanisms to our selected guidelines. In the manual analyses, three kinds of assessments are done to assess general features, policies, and privacy settings. In the static analysis, we evaluated apps’ privacy mechanisms by mobile security testing framework. To organize our overall set-up, we referenced research on apps security testing in different areas like mhealth, dating apps, fitness apps, followed by research on developers’ views of testing before releasing apps. We studied their methodology to have detailed insights on testing \cite{gilbert2011automating}, \cite{gao2014mobile}, \cite{zhang2017characterizing}, \cite{razaghpanah2018apps}. The following sections will briefly present our methodology.

\subsection{Data Sampling}
For performing an initial screening of possible applications of privacy, we collected 532 apps from the App Store. We used a variety of keywords to find relevant apps, including, ”Privacy,” ”Data Protection,” ”Hide,” ”Block,” ”Concealment,” ”Confidentiality,” ”Privateness,” ”Seclusion,” and ”Solitude.” We used certain specifics (e.g., language used, review count, ratings, category, prices and short description of their functionalities) to select the most suitable ones for assessment. Our total initial collection introduced us privacy apps of different categories, for example, photo and video privacy apps, VPN/ Wifi privacy, password manager, transmission and encryption, ad blocking, document/file privacy and others. \textbf{Figure 1 }provides a presentation of our collected apps’ data.

\begin{figure}
\centering
   \includegraphics[width=3.3in]{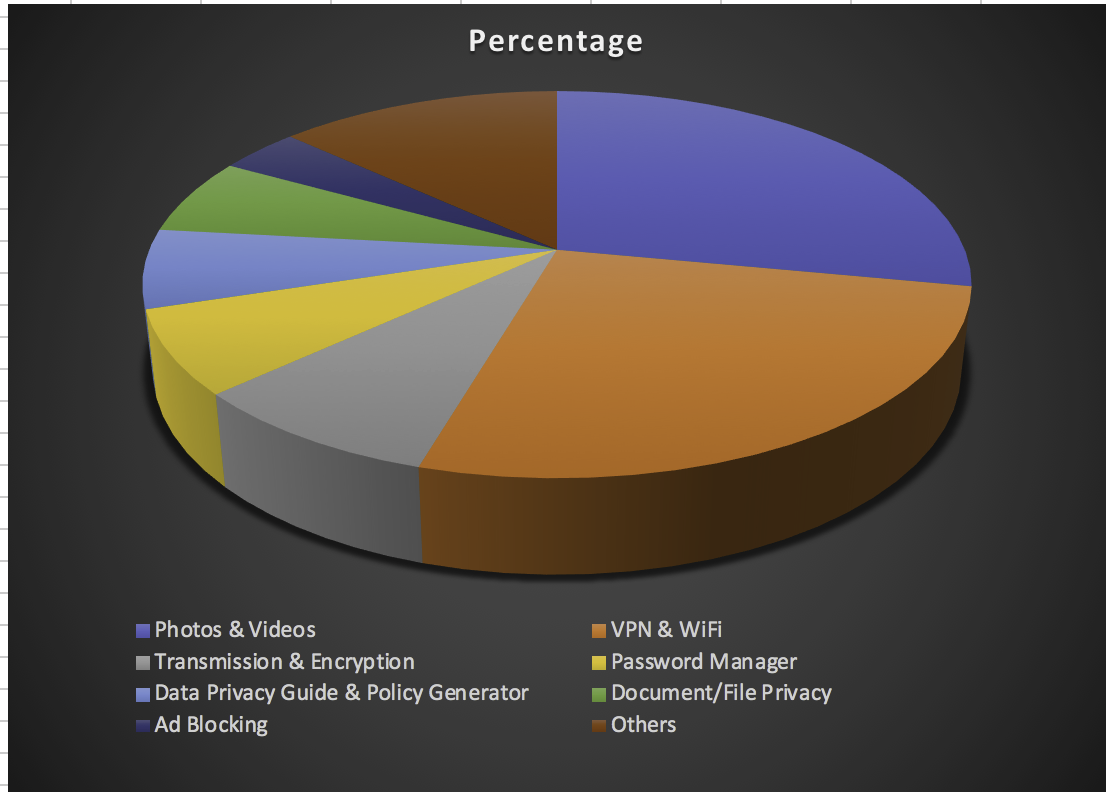}
\caption{Collected data of iOS privacy apps }
\label{fig:verticalcell}
\end{figure}

To identify, collect, and evaluate a manageable number of privacy apps, we determined our initial inclusion criteria. We chose those apps which are free, with instructions written in English, with a review count of more than 10798 (where 10798 is the mean of review counts of the total listed 532 apps), and with an overall rating higher than 3.5. By applying the inclusion criteria, (price, language, review count, rating) we ended up selecting 50 apps. Table 1 includes the inclusion criteria. However, 50 apps were still not a manageable number for in-depth analysis. Depending on our experiment, we choose to select apps that belong to six main areas of privacy: (i) Browsing privacy apps, (ii) VPN and WiFi Privacy apps, (iii) Photos and Videos Privacy apps, (iv) Password Privacy apps, (v) Ad blocking apps and (vi) Texting Privacy apps. We are utilizing mobile testing framework and automated tools for our complete set of analysis (static and dynamic). Based on the criteria mentioned  in table 1, we decided to select a total of 12 apps belonging to the 6 main categories, which are also open source. We selected our final apps for better comparison and contrast in regard to finding out the comprehensiveness of their privacy components. In our research, we have not directly mentioned name or other identifiers of analyzed apps. Therefore, we have decided to refer them as App 1 to App 12.

\begin{table}[ht]
\caption{Inclusion Criteria} 
\centering 
\begin{tabular}{c c c c} 
\hline 
Criterion & Type of Criterion  \\ [2.5ex] 
\hline 
Criterion 1 &  Content must be in English   \\ 
Criterion 2 & Apps must be free  \\
Criterion 3  &Apps review counts $\geq$ 10798  \\
Criterion 4 & Apps rating should be $\geq$ 3.5\\ [1ex] 
\hline 
\end{tabular}
\label{table:nonlin} 
\end{table}

\subsection{Assessment Methodology}
In this section, we designed our assessment strategy for evaluating the privacy mechanisms of the 12 collected categorized privacy apps. Particularly, we aimed to answer the following three main research questions:

What kind of privacy mechanism they are providing for protecting users’ information in respect to our selected guideline NIST?\\
What can be suggested to the app developer to consider while designing privacy for those apps?\\
To find the answers for those above questions, we designed the following assessment strategies to carry on our experiment.\\

\begin{enumerate}
\item  We first registered and installed all 12 selected apps from iOS App Store onto our experimental iOS device. Some of them required a valid Apple ID to create an app specific account during installation. We carefully read all the each apps general information and privacy mechanisms mentioned in the apps’ specific page and, if needed, their particular redirected websites for more concrete information. After installation, we again carefully went through all scopes and objectives of each app in terms of privacy settings, (for example, registration, lock, authentication, key management, homepage privacy tips, auto lock, auto-clear when logout) to accurately design the whole process of our evaluation for conducting manual analysis.
\item After a general characteristics evaluation, we inspected different types of permissions those apps automatically turned in from users’ devices. In addition, we inspected privacy policies from apps’ web pages to check if they mentioned those permissions that they asked for or if it automatically turned on in users’ devices. We also carefully noted those apps that did not have a detailed privacy policy on the App Store. As far as we know Apple requires all apps to have a privacy policy since October 3, 2018. They do not allow new apps to be accepted without a privacy policy, and existing apps will not be able to be updated without one. This requirement is strict, regardless of whether apps collect personal information from users or not.

\item After manual analysis, we completed an automated static code analysis utilizing framework MobSF (Mobile Security Framework). In this step, we have analyzed the IPA files on the framework to identify possible privacy and security vulnerabilities highlighted by different analyzers (binary analysis, security analysis) present in MobSF. We recorded our identified issues found on the static analyzer which are important to consider in building a comprehensive privacy design for apps, particularly privacy apps.

\item In this step, we performed a dynamic analysis for each app using Fiddler, a well-known web debugging proxy set-up. Every app was installed and tested to achieve the most accurate results of each app’s performance at the time of its dynamic analysis. We also studied the manual communication of each app and third parties in terms of data transmission and sharing. In the design scheme of our overall experiment, represented in Figure 2, this particular analysis is on the right side of the clients. This experiment’s main idea is to find out the data collected from users’ devices and their input through apps and how that information flows through different media and third parties.

\item After finding transmission and communication channels, we analyzed the web server configuration to determine the privacy and security level of HTTPS data transmission. For this examination and scoring of the web server configuration, we chose to use a free on-line service, SSL Labs, which enables remote testing of a web server’s security measures and privacy effectiveness against a host of current threats and vulnerabilities.

\item After having examined the apps with these three tools from different points of view, for example, quality of communication channel, transport layer security, generated apps permission, distribution of data among different media as static and manual analysis, we recorded and inspected privacy apps’ mechanisms in terms of security architecture, application structure, inter-process communication, application publishing and application attack surface.
\item In this step, we summarized our findings for each app. In addition, we compared and contrasted those findings with NIST and OWASP Mobile Security Testing Guide requirements for iOS apps privacy and security.

\item Finally, based on finding related to the latest NIST requirements and OWASP Mobile Security Testing Guide requirements, we performed a number of checks in order to make some recommendations and suggestions for app developers, as well as for the mobile apps testing Framework community, to reconsider necessary modifications and changes for further improvement.
\end{enumerate}

\begin{figure}
\centering
   \includegraphics[width=3.5in]{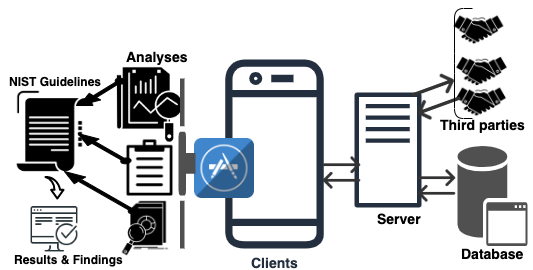}
\caption{ Scheme of the overall experiment }
\label{fig:verticalcell}
\end{figure}

Based on our overall analysis, our aim is to make sure that those apps are including privacy mechanisms as they claim in their apps’ information page, and therefore to make suggestions for developer to follow proper privacy guidelines. In addition, we also attempted to take the NIST guidelines in perspective to mark down if the mobile testing framework have available options to represent and examine privacy mechanisms comprehensively. Our findings may help in the development of a comprehensive privacy design for apps development, inspection and screening.

\section{Results}
Based on our methodology, we have drafted manual analysis report from a privacy policy and general settings points of view. After that, we have created a static and dynamic analysis results table for each app in different scenarios, for example, apps permission (camera, microphone, location, contacts, Bluetooth, storage, memory, etc.), privacy architectural requirement, communication and inter process communication security, and policy based on particular evaluation.

\subsection{Manual Analysis}
In manual analysis, we covered mainly three particular topics: Privacy policy, apps permission and general features related to privacy. Table2 represents the lists of existing characteristics to better understand the attributes to be factored in our evaluation as a form of score or percentage.

\begin{table} [htp]
\caption{Apps General and Privacy Features from Apps' Manual Analysis}
\centering
\footnotesize\setlength{\tabcolsep}{9.9pt}
\begin{tabular}{l@{\hspace{6pt}} *{13}{c}}
\toprule
\bfseries Features & \multicolumn{12}{c}{\bfseries App Number}\\ 
\cmidrule(l){2-13}
& 1 & 2 & 3 & 4 & 5 & 6 & 7 & 8 & 9 & 10 & 11 & 12 \\
\midrule
\bfseries General Features\\

\textbf{1.} Actionable
Application Lock
& \checkmark  & \checkmark & \checkmark &  \checkmark& \checkmark &  \checkmark& \checkmark &\checkmark  & \checkmark& \checkmark & \checkmark & \checkmark
\\
\textbf{2.} Option to Change \\
Preferred Language
& \checkmark  & \checkmark & \checkmark &  \checkmark& \checkmark &  \checkmark& \checkmark &\checkmark  & \checkmark& \checkmark & \checkmark & \checkmark
\\
\textbf{3.} Auto suggest and option\\
to give feedback & \xmark & \xmark&\xmark  &  \xmark& \xmark& \xmark& \checkmark& \xmark & \xmark & \checkmark& \xmark & \xmark
\\
\textbf{4.}Good Performance in \\
constrained environment 
& $\square$ &   \checkmark& $\square$ & $\square$ & $\square$& $\square$&$\square$&$\square$  &  \checkmark &  \checkmark &$\square$  &$\square$
\\
\textbf{5. }Storage Backup 
& $\square$ & $\square$ &  \checkmark& \xmark& \checkmark&\checkmark &$\square$  & $\square$  & $\square$  &\checkmark & \xmark& \checkmark
\\
\textbf{6. }Screen lock 
& $\square$  &  \checkmark&\checkmark  &  \xmark& \checkmark& \checkmark&\checkmark &  \checkmark&\checkmark  & \checkmark&\xmark  & \checkmark
\\
\textbf{7.} On-tap connect of configuring
&  \checkmark&  \xmark &  \xmark &  \xmark & \xmark & \xmark & \xmark & \xmark  & \xmark  &\checkmark &  \xmark & \xmark 
\\
\textbf{8.} Emergency Access feature by\\
other trusted party of the users
&  \xmark &   \xmark& \xmark  &   \xmark&\checkmark &  \xmark&  \xmark&   \xmark&   \xmark&  \xmark&   \xmark&  \xmark
\\
\textbf{9. }Simpler Interface
&  \checkmark&  \checkmark&  \checkmark& \checkmark & \checkmark& \checkmark& &  \checkmark&\checkmark  & \checkmark&\checkmark  &\checkmark
\\
\hline

\bfseries Privacy Features
&  &  &  &  & & & &  &  & &  & 
\\
\textbf{1.} Decoded Privacy Policy
&  \checkmark&  \checkmark&  \checkmark&  \checkmark& \checkmark& \xmark&\checkmark & \checkmark &  \checkmark& \checkmark&  \checkmark& \checkmark
\\
\textbf{2.} Well described Compliance:\\
with EU/US certificate
&  \checkmark&  \xmark&  \checkmark&  \xmark& \checkmark& \xmark& \checkmark&  \checkmark&  \checkmark&\checkmark & $\square$  & \checkmark
\\
\textbf{3.} Homepage Privacy Tips
&  \xmark&  \xmark& \xmark &  \xmark& \xmark& \xmark&\checkmark & \xmark &\xmark  &\xmark &\xmark  & \checkmark
\\
\textbf{4.} Doesn't store any \\
Personal Information
&  \checkmark&  \xmark&  \xmark&  \xmark&\xmark & \checkmark& \checkmark&  \checkmark& \xmark &\checkmark &  \xmark& \checkmark
\\
\textbf{5.} Doesn't Store IP address
&  \checkmark&  \xmark& $\square$    &  \xmark&\checkmark& \xmark&\checkmark &  \checkmark& $\square$   &\checkmark &\xmark  & \checkmark
\\
\textbf{6. }Doesn't hold sensitive\\
data for long 
& \checkmark &  \checkmark&  \xmark&  \checkmark& \checkmark& \checkmark&\checkmark &  \xmark&  \xmark&\checkmark &  \xmark&\xmark 
\\
\textbf{7.}Use local HTML5 to store \\
content information  
&  \xmark &  \checkmark&  \xmark &  \xmark & \checkmark& \xmark & \xmark &  \xmark & $\square$   & \xmark & $\square$   & \checkmark\\

\textbf{8.}Clear all data after use
&   \checkmark&   \checkmark& $\square$  &   \checkmark& \checkmark & \xmark &  \checkmark&  \xmark &\xmark   &  \checkmark&  \xmark & \checkmark\\
\textbf{9. }Smart Encryption 
&  \checkmark&  \checkmark& \checkmark &  \checkmark& \checkmark& \checkmark& \checkmark&  \checkmark&  \checkmark&\checkmark &\checkmark  &\checkmark \\
\textbf{10.} Two-Factor Authentication 
&  \xmark&  \xmark& \xmark &  \xmark&\xmark & \xmark&\xmark &  \xmark&  \xmark&\xmark &  \xmark&\xmark \\
\textbf{11.} Secure Communication
&  \checkmark&  \checkmark& \checkmark &  \checkmark&\checkmark& \checkmark&\checkmark &  \checkmark&\checkmark  & \checkmark&  \checkmark& \checkmark\\
\textbf{12.} Secure Registration and Screen lock 
& $\square$  &  \checkmark& \checkmark &  \checkmark&\checkmark & \checkmark&\checkmark &  \checkmark& \checkmark&\checkmark &  \checkmark&\checkmark \\
\textbf{13. }Security Audit and Grading
&\checkmark&  \xmark& \xmark &  \checkmark& \checkmark& \checkmark& \checkmark&  \xmark&\checkmark  & \xmark& \xmark & \checkmark
\\
\textbf{14.}Apps Cloud Security
& $\square$  & $\square$  &   \checkmark& $\square$  & $\square$ & $\square$ & $\square$ & $\square$  &$\square$   &$\square$  & $\square$  &$\square$  
\\
\textbf{15.} Privacy and Security Certification
& \xmark &  \xmark&\xmark  & \xmark & \checkmark &$\square$ & \xmark&  & \xmark & \xmark&  \xmark& \xmark
\\
\textbf{16. }Apps Complies with Privacy \\
standards and Principles
&$\square$  &  \xmark& \checkmark & \xmark&\checkmark & \xmark &\checkmark & \xmark & $\square$  &$\square$ & $\square$  & \checkmark
\\
\bottomrule
\addlinespace
\multicolumn{12}{l}{ $\square$ : not-mentioned}\\
\multicolumn{12}{l}{ \checkmark : Feature Presents in them}\\
\multicolumn{12}{l}{ \xmark: Feature doesn't present in them}
\end{tabular}
\end{table}

\subsubsection{Privacy Policy Inspection}
To maintain transparency on handing users’ personal information provided through apps and to give an outline of the terms of use and to outline the terms of use of the apps, a comprehensive and clear privacy policy is important. According to the iOS platform’s general criteria of publishing apps, they must have a privacy policy if the app involves collecting, processing or sharing personal data of users with other parties or even storing and processing user data for their own application development process \footnote{Apple developer documentation. https://developer.apple.com/documentation/}. Furthermore, an app needs a privacy policy even if it does not collect this kind of data itself but instead uses third-party tools like mobile analytics to collect and process data . iOS developers must read and agree to Apple Review Guidelines, a summary version based on their “Program License Agreement (PLA)” and other legal documents, in order to have their apps published on Apple App Store. iOS apps may get rejected if they do not add the URL to their privacy policy when they submit the app for review \footnote{App Store Review Guidelines. https://developer.apple.com/appstore/review/guidelines/privacy}.

Our evaluation of privacy policies started with the language inspection to audit if policies are in standard English. Each apps policy met the requirements. Keeping iOS platform restrictions and App Store review guidelines in mind, we studied the privacy policy of the 12 selected apps, and we found that 2 of those apps (app 10 and 12) did not have detailed privacy policy regarding their data collection and sharing, even though those apps are intended for privacy protection. Most of the apps did provide privacy policy and some with dubious statement in terms of malicious or unintentional data leakage, data breach and sharing. We also found some nontransparent policies regarding sharing users’ information with third parties, which somehow may meet the goals described in privacy policy. Another issue has been found in case of Cloud usage through apps for users’ data storage and backup, for example, apps providing secure vault for storing file, images, and videos. Perhaps statements on data minimization and retention can be more clear and transparent. However, excluding only one selected app, their privacy policies did not explain how the app’s Cloud security is performing and how privacy of that data is ensured, and most importantly how long that data will be there. Another significant point was privacy and security certification that has not been in practice in respect to our selected apps. Certification is a way to ensure credibility and build trust between users and organization. With the advancement of technology, most of the big tech giants and technological industries are concerned about valid product and service certifications which make them trustworthy to users/consumers. In the mobile apps scenario, this particular trend is missing. We found only 1 app out of 12 was maintaining privacy and security certification. While we started looking for standards followed in apps’ development stage and compliance with privacy standards and principles, we found only 3 apps out of 12 meet this requirement. Overall, we can conclude that manual analysis presents a lack of consideration on transparent statement on storage backups in privacy policy, app’s cloud security measurement, lack of practice in compliance with privacy standards and principles, and validation and audited by privacy and security certification in our evaluated 12 privacy apps.

\subsubsection{General and Privacy Features}
After thoroughly inspected privacy policy for each apps, we moved forward with apps general characteristics which includes both general and privacy features. For this stage of our work, we did not have any particular benchmark to collect specific information from apps specific web page or directly from apps while using those. We generally went through all the contents found in the Apps Store and screened through all the tabs and possible option during using those apps. We recorded the general features and the privacy and security components they mentioned and deployed to provide different functionalities and privacy preserving environment to their users. Some of these components are: users’ control over their data, smart encryption, decoded privacy policy, privacy grading based on tracker, and extra privacy features, personalized biometric locks options, security audits, data backups and so on. From the collected information on general and privacy features, we have noticed a bunch of handful functionalities. With keeping a note on performance of apps, we found lack of information regarding the performance of those apps in case of constrained environment, for example, with low bandwidth, high traffic and so on. As all of the apps require an internet connection, apps performance is a kind of information that might be taken under consideration. At least, there should have information about those apps’ performance statistics by some degree. In addition, these apps employ different privacy mechanisms, like encryption, which requires a high mathematical computation that can slow down apps’ performance in run-time. We noticed only very few apps mentioned their consistency and lack of backing in different environment. From our observation, we found Only 3 of the 12 apps had good performance in a constrained environment. \textbf{Table 2 } shows the overall recorded report from our 12 selected privacy apps for iOS mobile device on their basic requirements.

\subsubsection{Permission Analysis}
For preserving privacy of users’ information on apps’ platforms, one who is using those apps must grant permission to access personal information. This is sometimes useful if apps are using different information, for example, location, reminders to make our regular events convenient. While it is about privacy, users expect some level of privacy. If some permission request is not relevant of that apps’ purpose, it might seems a bit dubious when there is not enough explicitly mentioned description and policy in app’s web page. To analyze the permission requests of the selected privacy apps, we recorded the permissions listings from the IPA Manifest files of the 12 selected privacy apps. We categorized those permissions listings. Two main types of permission requests were  found: severe permission and normal permission. In our category of permission, we decided to include all of the permission requests mentioned in the iOS platform. Those are: photo, calendar, camera, location when in use, location always, location description, microphone, Apple music, bluetooth, and contacts. Figure 3 summarizes the requested permissions and represents the permission requests found by utilizing Mobile Security Framework.

\begin{figure}
\centering
   \includegraphics[width=3.4in]{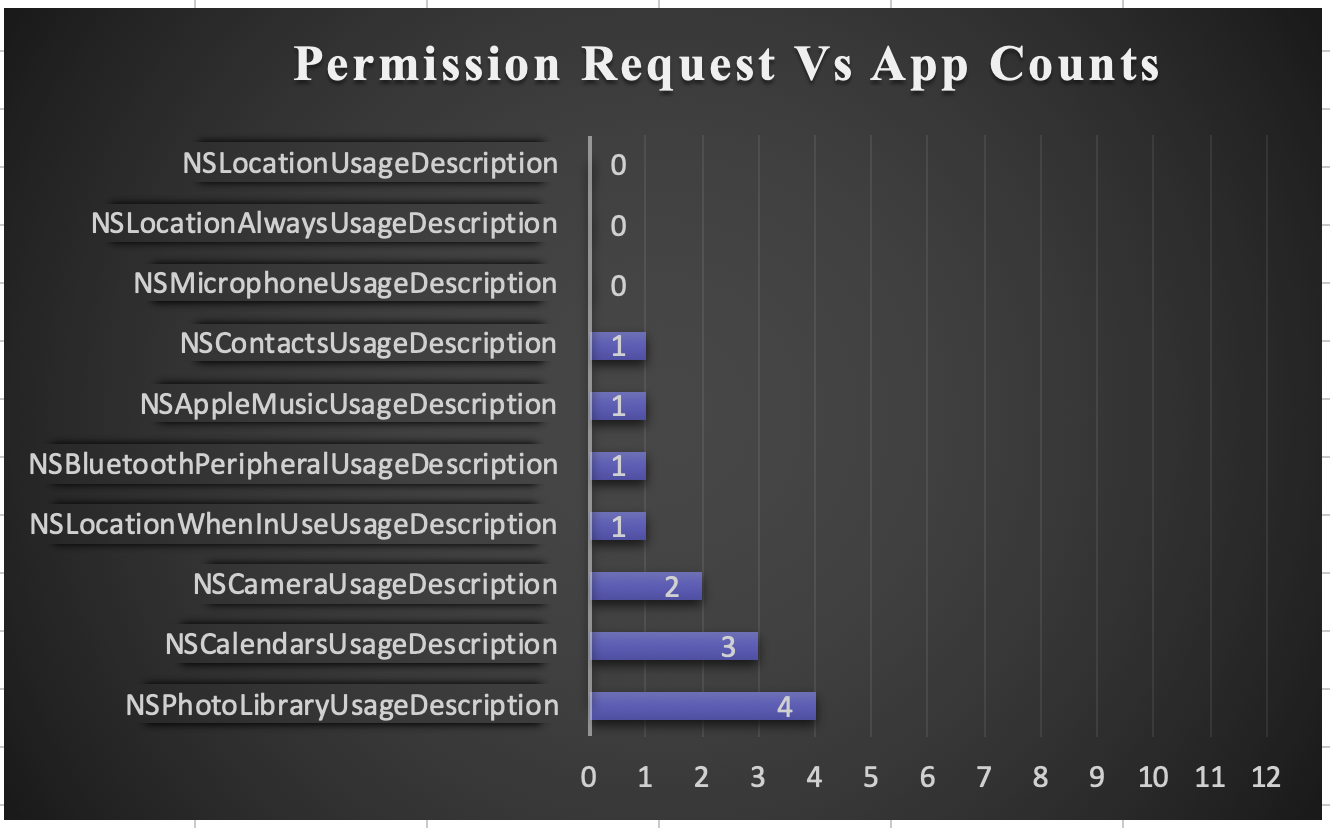}
\caption{ Summary of Permission requests}
\label{fig:verticalcell}
\end{figure}

From our experiment, we found that several apps have requested permission beyond their scope. For instance, four apps required access to the photo library and some of them are also requesting it without any obvious reason since we only had two apps in our list for the purpose of protecting photo and videos. While none of the apps in this study required any Bluetooth functionality or connectivity to a paired device, oddly, one of the applications requested for Bluetooth permission. According to our understanding, these additional permissions were requested for ad libraries, which exploit Bluetooth devices to track a user’s location. Notably, iOS required apps that performed scanning for hardware identifiers, like WiFi or Bluetooth, to request the location permission in direct approach with short alert message in their device to ask for granting permission for obtaining location information. Nevertheless, one of the studied apps requested permissions to access location and coarse location. Three of the apps requested calendars permission while two asked for camera permission. While only three apps needed access to the calendar, one more asked for access to the contacts list. One application requested access to Apple music permission. Remarkably, in most cases, permission requested by those apps do not have explicitly any functionality justifying such permission requests. Figure 3 provides a graphical representation of overall permission requests.
 
\subsection{Static Analysis}
For the static analysis, we evaluated the privacy and security and permission of the apps in detail. We examined each app independently using MobSF framework. The analysis revealed several privacy and security issues summarized in Table 3. Mainly we evaluated App transport security, static code analysis of IPA library and permissions. From 2016’s Data submission, Global survey of OWASP, the top 10 categories are focused towards Mobile applications \cite{jain2012addressing}. Insecure communication is one of top issues in mobile application since those applications are basically designed by client-server fashion while they transfer any information. We carry our mobile phone everywhere with us which make it more prone to network traffic where targeted attack is easier to perform.

\begin{table}[ht]
\footnotesize\setlength{\tabcolsep}{4.0pt}
\caption{Results of Static Code Analysis} 
\centering 
\begin{tabular}{c c c c} 
\hline\hline 
Code Analysis & Percentage  \\ [1.0ex] 
\hline 
Insecure API(s)&  67 
\\

\hline 
Insecure login Function &  83 
\\
\hline
Insecure Random Number Generator & 75  \\
\hline

No nemory protection mechanism (ASLR)\
& 25 \\
\hline
Insecure WebView Implementation & 50\\
\hline 
Insecure Implementation of SSL & 58\\
\hline

\end{tabular}
\label{table:nonlin} 
\end{table}

In most cases, mobile application use TLS/SSL at the time of authentication but not elsewhere. This can lead to data risk and session IDs. If adversaries get idea about users’ information, that attack can be more targeted in different form. Poor SSL setup can also facilitate phishing and MITM attacks \footnote{OWASP Mobile Top 10. Retrieved from https://owasp.org/www-project-mobile-top-10/
}. In the same way, insecure login function, insecure authorization and authentication have been creating a sparse attack vector by utilizing automated/bot attacks and custom-built tools. Here, we can mention our permission request analysis in our earlier section of this paper. If apps are performing the user’s roles or permissions to the backend system as part of a request, then it clearly shows that this app has issues with authorization. From our experiment by MobSF Framework, we have found 83\% of our smartphone privacy apps have insecure login function. For protecting system, there is a technology called ASLR (Address Space Layout Randomization) memory protection mechanism to help prevent attacks and shellcode from being successful. It esssentially protects system against buffer overflow attacks for which adversaries need to know where each part of the program is located in memory. ASLR randomly offsets the location of modules and certain in-memory structures to prevent this attack. 25\% of our selected apps do not have memory protection mechanism. On iOS platforms, apps are compiled with an Automatic Reference Counting (ARC) flag, which is a compiler that provides automatic memory management that reduces privacy and security risks against memory corruption and vulnerabilities. We found a larger portion of those apps were compiled with a Position Independent Executable (PIE) flag that address Space Layout Randomization (ASLR) which is also a memory protection mechanism to reduce memory corruption related vulnerabilities, while some apps were compiled with a Stack Smashing Protector (SSP) flag that builds protection against Stack Overflows attacks. From our static analysis, we also found some issues, such as insecure random number generators, which appear quite often, though not all qualified as significant. In most cases, random number generators usage was not necessarily related to security or privacy violations. Our framework evaluation also showed that 67\% of the apps have insecure API. Many of the apps do not use HTTPS while connecting, which is alarming, and raise several issues concerning WebViews components as we mentioned for API. 58\% apps have an insecure SSL connection and we examined it in further in our analysis by utilizing SSL lab.

In addition, we investigated more on Apps’ transport Security. An application needs to have ATS (Apps Transport Security) enabled to have secure communication. Insufficient transport security can lead to great impact on confidentiality of users’ information, more specifically, this is a wide area of research in respect to the practice of data transmission of eHealth apps \cite{tanusree2016feasibility}, as well as safety issues regarding data integrity \cite{muthing2017client}, \cite{rizzo2018babelview}. In case of iOS, this is a privacy feature that should be enabled by default when new apps are installed and enforces secure connections. ATS requires that all HTTP connections made with the URL Loading System—typically using the URLSession class-use HTTPS. It also includes extended security checks that conduct server trust evaluation prescribed by the Transport Layer Security (TLS) protocol and the main task of ATS to block connections that fail to meet minimum security specifications. It is also possible to remove and minimize requirements for communication with specific servers using the NSExceptionDomains key. From our app collections, we have found 3 apps have an insecure transport layer. Two of those three have their ATS disabled on "NSAllowsArbitraryLoads" for webviews. Another has ATS disabled for "Facebook.com", "Nomobileleads.com" and "Trymobilevpn.com" domains. Table 4 presents the status of ATS security level and issue found from our experiment.

\subsection{Dynamic Analysis}
During our dynamic analysis, we evaluated the apps in run-time using the selected apps on our devices to identify their communication channels. We considered their corresponding privacy and security attributes to find out if they transmit sensitive and personal data or any media that is not mentioned in their primary information over the Internet to unauthorized parties \cite{farshidi2016new}\cite{rahman2016pso}. We have identified some data types from previous researches and grouped accordingly, for example, multimedia, location that apps transmitted, registration and log-in data, e-mails, device I.D., search queries, OS information, and chat sessions. First, we examined the ATS (Apps Transport Security) mentioned in the record of static analysis to validate if potential insecure communication exists as found from static information. In iOS, App Transport Security (ATS) is used for security checks. When iOS operating system builds connections with NSURLConnection, NSURLSession and CFURL to public host-names, it requires some enforced requirement which is included in ATS. To maintain a secure transport, ATS should be enabled by default for applications running on iOS. When there are inconsistencies in enabling ATS for particular domains, it creates an insecure connection. Therefore, for any connection made to an IP address, unauthorized domain names of local are not protected with ATS \footnote{https://support.apple.com/guide/security/welcome/web}. Overall, the requirements of ATS are following: 

HTTP connections are not allowed. \\
X.509 Certificate has a SHA256 fingerprint and must be signed with at least a 2048-bit RSA key or a 256-bit Elliptic-Curve Cryptography (ECC) key\\

Transport Layer Security (TLS) version must be 1.2 or above and must support Perfect Forward Secrecy (PFS) through Elliptic Curve Diffie-Hellman Ephemeral (ECDHE) key exchange and AES-128 or AES-256 symmetric ciphers.\\
According to these requirements, we recorded the ATS overall status and causes for 12 selected privacy apps.
\begin{table}[ht]
\footnotesize\setlength{\tabcolsep}{3.0pt}
\caption{Apps Transport Security} 
\centering 
\begin{tabular}{c c c c} 
\hline\hline 
 Description& Issue & Status & App \\ [1.0ex] 
\hline  
ATS  is enabled & None & Secure & 1 \\
\hline
ATS is disabled on domain\\
'{'NSAllowsArbitraryLoads': True} & Present& Insecure & 2 \\
\hline
No insecure connections \\
configured ATS is enabled & None & Secure & 3 \\
\hline
No insecure connections \\
configured ATS is enabled & None & Secure & 4 \\
\hline
No insecure connections \\
configured ATS is enabled & None & Secure & 5 \\
\hline
No insecure connections \\
configured ATS is enabled & None & Secure & 6 \\
\hline 
No insecure connections \\
configured ATS is enabled & None & Secure & 7\\
\hline
ATS is disabled on domain \\
'{'NSAllowsArbitraryLoads': True}'that \\allow insecure communication and\\ loads for web views & Present & Insecure & 8\\
\hline
No insecure connections \\
configured ATS is enabled & None & Secure & 9 \\
\hline
No insecure connections \\
configured ATS is enabled & None & Secure & 10 \\
\hline
ATS is disabled domain 'facebook.com'\\
ATS is disabled on 'nomobileads.com'\\
ATS is disabled on'trymobilevpn.com
 &  Present & Insecure & 11\\
\hline
No insecure connections \\
configured ATS is enabled & None & Secure & 12 \\
\hline
\end{tabular}
\label{table:nonlin} 
\end{table} 

After the apps’ transport security analysis, we can see that some of the apps’ App Transport Security (ATS) is disabled on the domain ”NSAllowsArbitraryLoads:’ True. Deactivating ATS means allowing insecure communication with particular unauthorized servers, and therefore allowing insecure media loads for apps web views \cite{rizzo2018babelview}. Table 4 shows the result of ATS issues, status with description.

We also performed a privacy and security analysis while users interacted with those apps. We identified web traffic and data transmission nature by using Fiddler web debugging tool. This tool has a web filter and also a proxy. Normally the developer uses these tools for testing their application. Sometimes, with all changes in SSL, applications can be blocked, and it is difficult where this application is trying to go. In those cases, developers troubleshoot by using Fiddler to find the websites that create those exclusions.

\subsection{NIST 1800-21 and OWASP Assessment}
Apart from our previously described evaluations and different types of analysis, we also proceeded to an additional evaluation process in order to check and validate the requirements that the apps should meet in order to ensure data protection for users. We briefly studied NIST: NCCoE Component for Risk assessments, which include threat source, threat event, vulnerabilities, and some others. For our particular assessment and validation, we chose to delve into threat events and countermeasures to combat those. From NIST, we found 12 threat events, which is called T12 \cite{khidzir2010information}, and which includes: unauthorized access, credential theft, confidentiality and integrity, violation of privacy, network communication, unencrypted communication, device unlock code, storage vulnerabilities, device compromise, data loss, and unauthorized storage \cite{franklin2019mobile}. Using the NIST risk assessment, we created our list of countermeasures for seven primary threats and the requirements for those measures. Our selected measures for comprehensive privacy design included: architecture and design, data storage, cryptography, authentication, network communication, and code quality.

With briefly defined protection mechanisms and the threat definition of NIST 1800-21: Mobile Device Security requirements and mobile application testing tools and methods presented in OWASP, we performed an evaluation to compare and contrast. In this section, we reevaluated the results from previous sections and aligned them with the requirements of NIST and OWASP guidelines to find out what they were lacking \cite{franklin2019mobile}, \cite{mueller2017owasp}, \cite{morera2016security}. From our examination, we concluded that the requirements showed in Table 5 may help mobile apps privacy testing community to reconsider a few mechanisms and add a few features. We included the results for functional data protection requirements of NIST, such as privacy design and architecture, encryption, secure transmission or/and strong authentication. We present our results for the 12 apps accordingly. Table 5 shows the final results with NIST requirements.

\begin{table}[htp]
\caption{Selected Apps' Privacy and Security Checklist based on NIST and OWASP}
\centering
\footnotesize\setlength{\tabcolsep}{5.9pt}
\begin{tabular}{l@{\hspace{6pt}} *{12}{c}}
\toprule
\bfseries General Features & \multicolumn{12}{c}{\bfseries App Number} \\
\cmidrule(l){2-13}
& 1 & 2 & 3 & 4 & 5 & 6 & 7 & 8 & 9 & 10 & 11 & 12 \\
\midrule
\bfseries Architecture and design\\
\hline
1. Security controls at Client side 
& \checkmark & \checkmark & \checkmark & \checkmark & \xmark & \xmark& \checkmark& \xmark & \checkmark & \xmark & \xmark & \checkmark 
\\
\hline
2. High-level privacy architecture
& \checkmark & \xmark & \xmark & \xmark & \checkmark & \xmark& \checkmark & \checkmark & \checkmark & \checkmark & \xmark & \checkmark 
\\
\hline
3. Clear context of data sensitivity
& \checkmark & \checkmark & \xmark & \xmark& \checkmark & \checkmark & \checkmark & \xmark & \checkmark & \checkmark & \xmark & \checkmark
\\
\hline
4. Defined security functionality of content
& \checkmark & \xmark & \checkmark & \xmark & \checkmark & \xmark & \checkmark & \checkmark & \checkmark & \checkmark & \xmark & \checkmark 
\\
\hline
5. Threat model,countermeasures
& \xmark & \xmark & \xmark & \xmark& \xmark & \xmark& \checkmark & \xmark & \checkmark & \checkmark & \xmark & \checkmark 
\\
\hline
6. Explicit key-management standard (NIST SP 800-57)

& \checkmark & \checkmark & \checkmark & \checkmark & \checkmark & \checkmark & \checkmark& \checkmark & \xmark & \checkmark &  \xmark & \checkmark 
\\
\hline
7. Mechanism for enforcing updates 
& \xmark & \xmark & \xmark & \xmark & \checkmark & \xmark & \xmark & \checkmark & \checkmark & \checkmark & \xmark & \xmark
\\
\hline
8. Privacy addressing in soft development life cycle
& \xmark & \xmark & \xmark & \xmark & \xmark& \xmark& \checkmark & \xmark & \xmark & \checkmark & \xmark & \checkmark
\\
\hline
\bfseries Data Storage and Privacy
& \checkmark & \checkmark & \checkmark & \xmark & \checkmark & \checkmark & \checkmark & \checkmark & \xmark & \checkmark & \xmark& \checkmark
\\
\hline
1. Cryptographic keys for sensitive
Info
& \checkmark & \checkmark & \xmark & \checkmark & \checkmark & \checkmark & \checkmark & \checkmark & \checkmark & \checkmark & \xmark & \xmark 
\\
\hline
2. Doesn't storage data Outside container
& \checkmark & \xmark & \xmark & \xmark & \xmark & \checkmark & \checkmark & \checkmark & \checkmark & \checkmark & \xmark & \checkmark 
\\
\hline
3. Cryptographic keys for sensitive
& \checkmark & \checkmark &  \xmark & \checkmark& \checkmark& \checkmark & \checkmark & \checkmark& \checkmark & \xmark & \xmark & \checkmark 
\\
\hline
4. Doesn't share data with 3rd parties
& \xmark & \xmark& \xmark & \xmark & \xmark & \xmark & \checkmark & \checkmark & \checkmark & \xmark & \xmark & \checkmark 
\\
\hline
5. Sensitive data exposition via IPC mechanisms.
& \xmark& \xmark & \checkmark & \xmark & \checkmark & \xmark & \checkmark & \xmark & \checkmark & \checkmark & \xmark & \xmark
\\
\hline
6. Passwords or pins exposition through\\
user interface
& \checkmark & \checkmark & \checkmark & \xmark & \xmark & \checkmark & \checkmark & \checkmark& \checkmark & \checkmark & \xmark & \checkmark 
\\
\hline
7. Doesn't hold sensitive data for long 
& \checkmark & \checkmark& \xmark & \xmark& \xmark & \checkmark & \checkmark & \checkmark & \checkmark & \checkmark & \xmark & \checkmark
\\
\hline
8. App have education materials for users 
& \checkmark & \xmark & \xmark & \xmark & \xmark & \xmark & \checkmark & \checkmark & \checkmark& \checkmark & \xmark & \checkmark  
\\
\hline
\bfseries Cryptography  \\
\hline
1. Symmetric cryptography with hardcoded \\
keys as a sole method of encryption.
& \checkmark & \checkmark & \checkmark & \checkmark & \checkmark & \checkmark & \checkmark & \checkmark & \checkmark & \checkmark & \xmark & \checkmark 
\\
\hline
2. Implementations of cryptographic\\
primitives.
& \checkmark & \checkmark & \checkmark & \checkmark & \checkmark & \checkmark & \checkmark & \checkmark & \xmark & \xmark& \xmark & \xmark
\\
\hline
3. Doesn't re-use the same cryptographic\\
key for multiple purposes
& \checkmark & \xmark & \checkmark & \xmark & \checkmark & \xmark & \checkmark & \checkmark & \xmark & \xmark & \xmark & \checkmark 
\\
\hline
4. All random values are generated using\\
secure random number generator
& \xmark & \xmark & \xmark& \xmark & \checkmark& \checkmark & \checkmark& \xmark& \checkmark & \checkmark & \xmark & \checkmark 
\\
\hline
\bfseries Authentication and Session Management  \\
\hline
1. Stateful session management
& \checkmark & \xmark & \xmark & \checkmark & \checkmark & \checkmark & \checkmark & \checkmark & \checkmark & \checkmark & \xmark & \xmark
\\
\hline
2. Provides access to a remote service
& \checkmark & \xmark & \xmark & \xmark & \checkmark & \checkmark & \checkmark & \checkmark & \xmark& \checkmark & \xmark & \checkmark
\\
\hline
3. Terminates existing session on logout
& \checkmark & \checkmark & \checkmark & \xmark & \checkmark & \checkmark & \checkmark & \checkmark & \xmark & \xmark & \xmark & \checkmark
\\
\hline
4. Biometric authentication
& \xmark & \xmark & \checkmark & \checkmark& \checkmark& \xmark & \xmark & \checkmark & \checkmark & \checkmark & \xmark& \checkmark
\\
\hline
5. 2FA authentication
& \xmark & \xmark & \xmark & \xmark & \checkmark & \xmark& \xmark& \checkmark & \checkmark & \checkmark &  \checkmark & \checkmark 
\\
\hline
\bfseries Network Communication  \\
\hline
1.Data encrypted on network using TLS
& \checkmark& \checkmark & \xmark & \checkmark & \checkmark & \xmark & \checkmark & \checkmark & \checkmark & \checkmark & \xmark & \xmark
\\
\hline
2.Verifies X.509 certificate of \\
remote endpoint 
& \xmark& \xmark & \xmark& \xmark & \checkmark& \checkmark & \xmark& \checkmark & \checkmark & \checkmark & \xmark & \checkmark
\\
\hline
3. Doesn't rely on a single insecure\\ communication channel 
& \checkmark & \xmark & \checkmark & \checkmark& \checkmark & \xmark & \checkmark& \checkmark & \checkmark & \checkmark & \xmark & \xmark
\\
\hline
4. depends on up-to-date connectivity and\\ security libraries
& \checkmark & \xmark & \checkmark & \xmark & \xmark & \xmark & \xmark & \checkmark & \checkmark & \checkmark & \xmark & \checkmark 
\\
\hline
\bfseries Platform Interaction  \\
\hline

1.Requests minimum set of permissions
& \checkmark & \checkmark & \checkmark & \xmark & \checkmark & \checkmark & \checkmark& \checkmark & \checkmark & \checkmark & \xmark & \checkmark 
\\
\hline
2.Doesn't export sensitive functionality\\
via custom URL scheme
& \checkmark & \checkmark & \checkmark & \checkmark & \checkmark & \checkmark & \checkmark& \checkmark & \checkmark & \checkmark & \xmark & \checkmark 
\\
\hline
3.Doesn't export sensitive functionality\\
by IPC facilities
& \checkmark & \checkmark & \checkmark & \xmark & \xmark & \xmark & \checkmark & \checkmark & \checkmark & \xmark& \xmark & \checkmark 
\\
\hline
4. JavaScript is disabled in WebViews
& \checkmark & \checkmark & \checkmark & \checkmark & \checkmark & \checkmark & \checkmark & \checkmark & \checkmark & \xmark & \xmark & \checkmark 
\\
\hline
\bfseries Code Quality and Build Settings  \\
\hline
1.Signed and provisioned with valid\\
certificate
& \checkmark & \checkmark & \xmark & \xmark & \xmark & \xmark & \xmark & \checkmark & \checkmark & \checkmark& \xmark & \checkmark 
\\
\hline
2.In unmanaged code, memory is allocated, \\
freed and used securely
& \xmark & \xmark & \xmark & \checkmark & \checkmark & \xmark & \checkmark & \checkmark & \checkmark & \checkmark & \checkmark & \checkmark 
\\
\hline
3. Free security features offered \\ 
by the toolchain

& \checkmark & \checkmark & \checkmark & \checkmark & \checkmark & \xmark & \checkmark & \checkmark & \checkmark & \xmark & \xmark& \xmark 
\\
\hline

\bottomrule
\addlinespace
\multicolumn{12}{l}{ \checkmark: Features Presents in them}\\
\multicolumn{12}{l}{ \xmark: Feature doesn't present in them}
\end{tabular}
\end{table}

\section{Discussion}
Smartphone app developers have many opportunities to create, collect, and share data about their users. Users might be unaware of the data collection, particularly when it is performed by sensors running in the background. Although this data can be used to create a better customer experience, it can also be used to make unsettling inferences about the app users’ habits, lifestyles, locations, and other information \cite{balebako2014improving}. These are creating privacy risks. While these are privacy apps, the issues can be more broadly serious. Privacy apps are mainly based on preserving users’ privacy. However, there still needs to be analyses to validate these kinds of apps if they are including proper privacy mechanisms and following privacy specific guidelines. Despite the privacy risks of this unprecedented access to data, many app developers are neither privacy nor security experts. The app development workforce is highly fragmented. Apps are often initially developed by independent developers or small start-ups, as opposed to large, established corporations. This fragmented market means that many independent developers with limited privacy experience and fewer personnel and resources are making privacy decisions about their users’ sensitive data and privacy mechanisms. To examine how app developers make privacy decisions, including what kind of privacy guidelines and mechanisms they are using, our study sheds light a on the lack of privacy mechanisms in existing privacy apps (the 12 selected privacy apps).

\begin{figure}
\centering
   \includegraphics[width=3.0in]{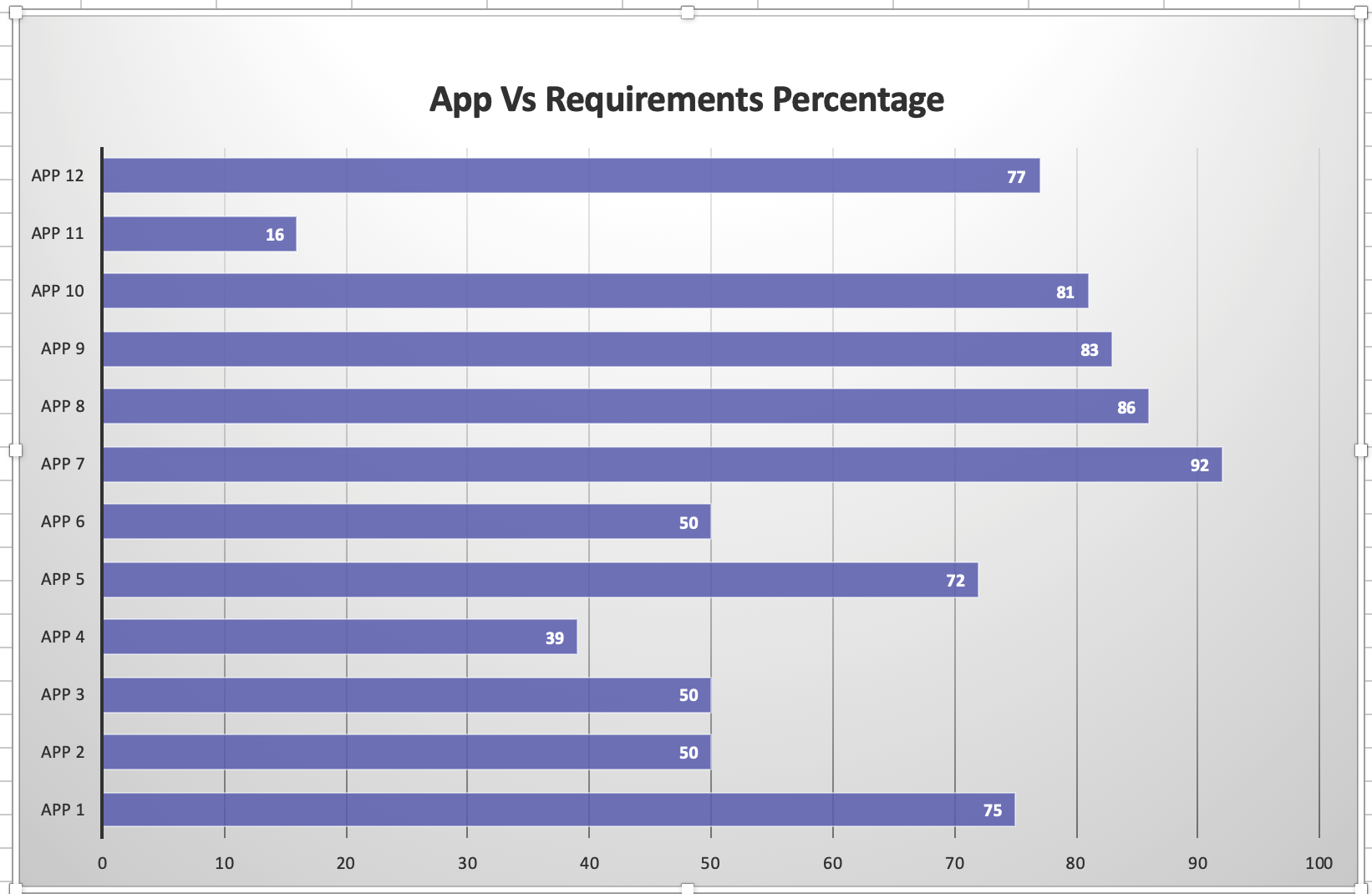}
\caption{Final apps presentation with percentage of requirements}
\label{fig:verticalcell}
\end{figure}
US government agencies, trade associations, and advocacy groups have released app developer guidelines, and different documents describe how developers can protect user privacy. For example, the Federal Trade Commission released a staff report on mobile privacy disclosures, and the California Attorney General provided recommendations for privacy in the mobile ecosystem. To synthesize these different documents, the International Association of Privacy Professionals (IAPP) created a Web tool that lets readers access, search, and compare 10 different privacy guidelines from the US, Australia, and Europe \footnote{Mobile Privacy Disclosures. Building Trust Through Transparency. FTC Staff Report | February 2013. Retrieved from https://www.ftc.gov/}. We examined the documents in the IAPP Mobile App Privacy Tool. For more technical guidance for mobile apps developers, NIST also published a comprehensive guideline with privacy threats and requirements. The guidelines often overlap, making it feasible to comply with most or all procedures. For our evaluation process, we have used the NIST guideline to compare our experimental results.

For our experiment and analyses, we selected 12 privacy apps and to avoid legal issues, we chose to present them as App 1-12, rather than revealing their names. We performed three forms of analyses: manual, static and dynamic. From the very first analysis, we obtained apps’ general and privacy features provided and claimed apps’ specific sites (Table 2). We were able to see almost perfect privacy features that have been used to develop their privacy apps. All of the apps include smart encryption, extra privacy registration PIN, secure communication, and secure data management, which are the most important criteria to provide privacy to users. In this stage, we have also done permission analyses with each apps’ IPA file to investigate if they are allowing any dangerous permission or not. Unlike android apps, they are not giving a lot of dangerous permission \cite{papageorgiou2018security}. Still, a few of the apps are requiring dangerous permissions like photo (4 apps), calendars (3 apps), camera (2 apps), location (1 app), Bluetooth (1 app), music (1 app), contacts (1 app). From the static analysis, we recorded a deficiency in coding in terms of log in, random number generator, SQLite database, etc. We have also found insecure connections in apps’ transport security for 3 apps from them and those are severe where even ATS is not enabled. As privacy apps, we expected not to have any third party involvement. However, almost all the apps have third parties involved in users’ data. Though the number are less than any android or any other apps in iOS, this is still concerning in terms of privacy apps. We have identified HTTP connection to 3rd parties per SSL with grades where there exist some grades that are not expected for privacy apps.

At the end of all experiments, we compared our results with NIST and OWASP guidelines to validate our findings. We have observed a huge gap in what we have got in our initial manual analysis in apps specific sites and what we got from our analyses result. There is scope for improvement of these apps from the permission requirement, HTTP connection, Apps Transport security point of view. As those apps are developed by multiple developers, they might follow different privacy guideline and requirements. In this case, we can suggest them to utilize comprehensive apps privacy guideline

\section{Conclusion}
Privacy apps includes mechanism that can ensure a certain level of data protection among cellphone. To validate those apps credibility, there needs a concrete guideline and certification process in their development life cycle. Nowadays, it is common in a wide marketplace of apps for developers to follow different self-regulatory guidelines. Even though the availability of development tools and comparably easy ways to publish apps in the market, amateurs do not even follow guidelines to ensure criteria of privacy mechanisms which eventually make users vulnerable. In particularly, for privacy apps development, it is crucial to follow widely recognized privacy guidelines. Through our research, we assessed the current state of practices for developing privacy apps. Our analyses of selected privacy apps identified their lacking in security and privacy mechanisms that can lead users to great vulnerabilities in case of data transfer, duration of data storage, inter-process communication, apps transport security, and apps permission requests, Our comparison and contrast with our findings based on NIST guidelines and another operational guideline (OWASP) requirements and controls highlighted the shortcomings of those apps. In addition, it also showed the lacking in the comprehensiveness in the framework that we utilized for our experiment which do not measure up to privacy mechanisms very well since those are basically developed to lean solely towards security aspects. Our overall results may help developers with further improvement of privacy apps’ design and development. For easily implementable privacy standards for apps development, a comprehensive privacy design with proper requirements will be a good start.
\bibliographystyle{unsrt}  
\bibliography{references}  



\end{document}